\documentclass[twocolumn,aps,prd,amsmath,amssymb,nofootinbib]{revtex4-1}\newcommand{\cw}{\columnwidth}\newcommand{\preprintnumber}{\hfill MIT-CTP 4499\,\,\,\,\maketitle}

\usepackage{graphicx,bm}
\usepackage{color,soul}

\newcommand{\beq}{\begin{equation}}
\newcommand{\bea}{\begin{eqnarray}}
\newcommand{\eeq}{\end{equation}}
\newcommand{\eea}{\end{eqnarray}}
\newcommand{\amp}{&}
\newcommand{\la}{\langle}
\newcommand{\ra}{\rangle}
\newcommand{\mpl}{M_{Pl}}
\newcommand{\N}{\mathcal{N}}
\newcommand{\Nw}{N_w}
\newcommand{\dt}{\delta t}
\newcommand{\xv}{{\bf x}}
\newcommand{\yv}{{\bf y}}

\newcommand{\ov}{{\bf 0}}
\newcommand{\kv}{{\bf k}}
\newcommand{\mphi}{m_0}
\newcommand{\mpsi}{m_\psi}
\newcommand{\phii}{\Phi}

\begin{document}

\title{A Density Spike on Astrophysical Scales from an $\N$-Field Waterfall Transition}

\author{Illan F.~Halpern$^{1,2,*}$, Mark P.~Hertzberg$^{1,\dagger}$, Matthew A.~Joss$^{1,\ddagger}$, Evangelos I.~Sfakianakis$^{1,3,\mathsection}$}
\affiliation{$^1$Center for Theoretical Physics and Dept.~of Physics,\\ Massachusetts Institute of Technology, Cambridge, MA 02139, USA
\\
$^2$Perimeter Institute for Theoretical Physics, Waterloo, ON N2L 2Y5, Canada
\\
$^3$Dept.~of Physics, University of Illinois at Urbana-Champaign, Urbana, IL 61801, USA}

\date{\today}

\begin{abstract}
Hybrid inflation models are especially interesting as they lead to a spike in the density power spectrum on small scales, compared to the CMB, while also satisfying current bounds on tensor modes. Here we study hybrid inflation with $\N$ waterfall fields sharing a global $SO(\N)$ symmetry. The inclusion of many waterfall fields has the obvious advantage of avoiding topologically stable defects for $\N>3$. We find that it also has another advantage: it is easier to engineer models that can simultaneously (i) be compatible with constraints on the primordial spectral index, which tends to otherwise disfavor hybrid models, and (ii) produce a spike on astrophysically large length scales. The latter may have significant consequences, possibly seeding the formation of astrophysically large black holes. We calculate correlation functions of the time-delay, a measure of density perturbations, produced by the waterfall fields, as a convergent power series in both $1/\N$ and the field's correlation function $\Delta(x)$. We show that for large $\N$, the two-point function is $\la \dt(\xv)\,\dt(\ov)\ra \propto \Delta^2(|\xv|)/\N$ and the three-point function is $\la\dt(\xv)\,\dt(\yv)\,\dt(\ov) \ra \propto \Delta(|\xv-\yv|)\Delta(|\xv|)\Delta(|\yv|)/\N^2$. In accordance with the central limit theorem, the density perturbations on the scale of the spike are Gaussian for large $\N$ and non-Gaussian for small $\N$. 
\end{abstract}

\preprintnumber

\newpage
\tableofcontents

\let\thefootnote\relax\footnotetext{$^*$Electronic address: {\tt ihalpern@perimeterinstitute.ca}\\$^\dagger$Electronic address: {\tt mphertz@mit.edu}\\
$^\ddagger$Electronic address: {\tt mattjoss@mit.edu}\\$^\mathsection$Electronic address: {\tt esfaki@illinois.edu}}

\section{Introduction}

Inflation, a phase of accelerated expansion in the very early universe thought to be driven by one or several scalar fields, is our paradigm of early universe cosmology \cite{Guth:1980zm,Linde:1981mu,Albrecht:1982wi,inflation}. It naturally explains the large scale homogeneity, isotropy, and flatness of the universe. Moreover, the basic predictions of even the simplest single field slow roll models, giving approximate scale invariance and small non-Gaussianity in the $\sim 10^{-5}$ level departures from homogeneity and isotropy, are in excellent agreement with recent CMB data \cite{Hinshaw:2012aka,Ade:2013uln, Martin:2013nzq} and large scale structure. 

While the basic paradigm of inflation is in excellent shape, no single model stands clearly preferred. Instead the literature abounds with various models motivated by different considerations, such as string moduli, supergravity, branes, ghosts, Standard Model, etc \cite{ModLinde, ModLiddle,ModDvali, ModMcAllister, ModNima, ModKachru,higgs,Greenwood:2012aj,particlephys,Martin:2014vha,Kaiser:2013sna,Hertzberg:2014aha}. While the incoming data is at such an impressive level that it can discriminate between various models and rule out many, such as models that overpredict non-Gaussianity, it is not clear if the data will ever reveal one model alone. An important way to make progress is to disfavor models based on theoretical grounds (such as issues of unitarity violation, acausality, etc) and to find a model that is able to account for phenomena in the universe lacking an alternate explanation. It is conceivable that some version of the so-called ``hybrid inflation" model may account for astrophysical phenomena, for reasons we shall come to.

The hybrid inflation model, originally proposed by Linde \cite{linde}, requires at least two fields. One of the fields is light and another of the fields is heavy (in Hubble units). The light field, called the ``timer", is at early times slowly rolling down a potential hill and generates the almost scale invariant spectrum of fluctuations observed in the CMB and in large scale structure. The heavy field, called the ``waterfall" field, has an effective mass that is time-dependent and controlled by the value of the timer field. The waterfall field is originally trapped at a minimum of its potential, but as its effective mass-squared becomes negative, a tachyonic instability follows, leading to the end of inflation; an illustration is given in Figure \ref{fig:Potential_Diagram}. 
The name ``hybrid inflation'' comes from the fact that this model is a sort of hybrid between a chaotic inflation model and a symmetry breaking inflation model. 

As originally discussed in Refs.~\cite{sol1,sol2}, one of the most fascinating features of hybrid models is that the tachyonic behavior of the waterfall field leads to a sharp ``spike" in the density power spectrum. This could seed primordial black holes \cite{lindeBH,carr1,carr2,carr3,carr4,Clesse:2015wea}.  
For generic parameters, the length scale associated with this spike is typically very small.
However, if one could find a parameter regime where the waterfall phase were to be prolonged, lasting for many e-foldings, say $\Nw\sim 30-40$, then this would lead to a spike in the density perturbations on astrophysically large scales (but smaller than CMB scales).
This may help to account for phenomena such as supermassive black holes or dark matter, etc. 
Of course the details of all this requires a very careful examination of the spectrum of density perturbations, including observational constraints.

Ordinarily the spectrum of density perturbations in a given model of inflation is obtained by decomposing the inflaton field into a homogeneous part plus a small inhomogeneous perturbation. However, for the waterfall fields of hybrid inflation, this approach fails since classically the waterfall field would stay forever at the top of a ridge in its potential. It is the quantum perturbations themselves that lead to a non-trivial evolution of the waterfall field, and therefore the quantum perturbations cannot be treated as small. Several approximations have been used to deal with this problem \cite{sol1,sol2, lyth1, lyth2, lyth3, stoch, sasaki,kodama,wands,kristin,son,Levasseur:2013tja, Clesse:2010iz, Clesse:2013jra}. Here we follow the approach presented by one of us recently in Ref.~\cite{sfak}, where a free field time-delay method was used, providing accurate numerical results. 

In this paper, we generalize the method of Ref.~\cite{sfak} to a model with $\N$ waterfall fields sharing a global $SO(\N)$ symmetry. 
A model of many fields may be natural in various microscopic constructions, such as grand unified models, string models, etc.
But apart from generalizing Ref.~\cite{sfak} to $\N$ fields, we also go much further in our analysis: we derive explicit analytical results for several correlation functions of the so-called time-delay. We formulate a convergent series expansion in powers of $1/\N$ and the field's correlation function $\Delta(x)$. We find all terms in the series to obtain the two-point correlation function of the time-delay for any $\N$. We also obtain the leading order behavior at large $\N$ for the three-point function time-delay, which provides a measure of non-Gaussianity. We find that the non-Gaussianity is appreciable for small $\N$ and suppressed for large $\N$.

We also analyze in detail constraints on hybrid inflation models. We comment on how multiple fields avoids topological defects, which is a serious problem for low $\N$ models. However, the most severe constraint on hybrid models comes from the requirement to obtain the observed spectral index $n_s$. We show that at large $\N$, it is easier to engineer models that can fit the observed $n_s$, while also allowing for a prolonged waterfall phase.
This means that large $\N$ models provide the most plausible way for the spike to appear on astrophysically large scales and be compatible with other constraints. 

With regards to tensor modes,
the confirmation of the recent detection and amplitude of primordial gravitational waves \cite{bicep} depends largely on understanding dust foregrounds \cite{bicep_planck}. The final answer to the existence of primordial tensor modes will have to be given by future experiments, since a recent joint analysis of current data \cite{Ade:2015tva} does not provide a clear detection.
So we will only briefly discuss possibilities of obtaining observable tensor modes from hybrid inflation in Section \ref{Constraints}.

The organization of our paper is as follows:
In Section \ref{model} we present our hybrid inflation model and discuss our approximations. 
In Section \ref{timedelay} we present the time-delay formalism, adapting the method of Ref.~\cite{sfak} to $\N$ fields.
In section \ref{correlation} we derive a series expansion for the two-point function, we derive the leading order behavior of the three-point function, and we derive results in $k$-space.
In Section \ref{Constraints} we present constraints on hybrid models, emphasizing the role that $\N$ plays.
In Section \ref{discussion} we discuss and conclude.
Finally, in the Appendices we present further analytical results.

\section{$\N$ Field Model}\label{model}

The model consists of two types of fields:
The timer field $\psi$ that drives the first slow-roll inflation phase, and the waterfall field $\phi$ that becomes tachyonic during the second phase causing inflation to end.
In many hybrid models, $\phi$ is comprised of two components, a complex field, but here we allow for $\N$ real components $\phi_i$. We assume the components share a global $SO(\N)$ symmetry, and so it is convenient to organize them into a vector
\beq
\vec\phi=\{\phi_1,\phi_2,\ldots,\phi_{\N}\}.
\eeq
For the special case $\N=2$, this can be organized into a complex field by writing $\phi_{complex}=(\phi_1+i\,\phi_2)/\sqrt{2}$.

The dynamics is governed by the standard two derivative action for the scalar fields $\psi,\vec\phi$ minimally coupled to gravity as follows (signature $+---$)
\bea
S=\int\! d^4x \sqrt{-g} \Bigg{[}{1\over16\pi G}R\amp+\amp \frac{1}{2} g^{\mu \nu}\partial_\mu \psi \partial_\nu \psi  \nonumber\\
\amp\amp\!\!\!\!\!\!\!\!\!\!\!\!\!\!\!\!\!\!\!\!\!\!\!\!\!\! + \frac{1}{2} g^{\mu \nu}\partial_\mu \vec\phi \cdot\partial_\nu \vec\phi- V(\psi,\phi) \Bigg{]}.\,\,\,\,
\eea
The potential $V$ is given by a sum of terms: $V_0$ providing false vacuum energy, $V_\psi(\psi)$ governing the timer field, $V_\phi(\phi)$ governing the waterfall field, and $V_{int}(\psi,\phi)$ governing their mutual interaction, i.e.,
\beq
V(\psi,\phi)=V_0+V_\psi(\psi)+V_\phi(\phi)+V_{int}(\psi,\phi).
\eeq
During inflation we assume that the constant $V_0$ dominates all other terms. 

The timer field potential $V_\psi$ and the waterfall field potential $V_\phi$ can in general be complicated. In general, they are allowed to be non-polynomial functions as part of some low energy effective field theory, possibly from supergravity or other microscopic theories. For our purposes, it is enough to assume an extremum at $\psi=\phi=0$ and expand the potentials around this extremum as follows:
\bea
&&V_\psi(\psi)={1\over2}\mpsi^2 \psi^2+\ldots\\
&&V_\phi(\phi)=-{1\over2}\mphi^2\, \vec\phi\cdot\vec\phi+\ldots
\eea
The timer field is assumed to be light $\mpsi<H$ and the waterfall field is assumed to be heavy $\mphi>H$, where $H$ is the Hubble parameter.
In the original hybrid model, this quadratic term for $V_\psi$ was assumed to be the entire potential. This model leads to a spectrum with a spectral index $n_s>1$ and is observationally ruled out. Instead we need higher order terms, indicated by the dots, to fix this problem. This also places constraints on the mass of the timer field $\mpsi$, which has important consequences. We discuss these issues in detail in Section \ref{Constraints}.

As indicated by the negative mass-squared, the waterfall field is tachyonic around $\phi=\psi=0$. This obviously cannot be the entire potential because then the potential would be unbounded from below. Instead there must be higher order terms that stabilize the potential with a global minimum near $V\approx0$ (effectively setting the late-time cosmological constant).

\begin{figure}[t]
\includegraphics[width=\cw]{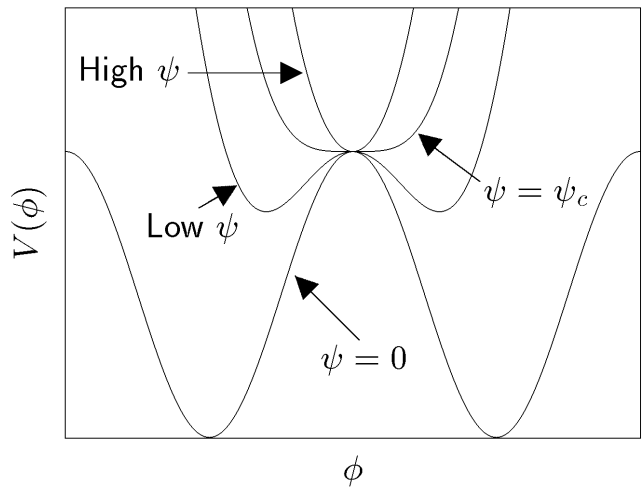}
\caption{An illustration of the evolution of the effective potential for the waterfall field $\phi$ as the timer field $\psi$ evolves from ``high" values at early times, to $\psi=\psi_c$, and finally to ``low" values at late times. In the process, the effective mass-squared of $\phi$ evolves from positive, to zero, to negative (tachyonic).}
\label{fig:Potential_Diagram}
\end{figure}

The key to hybrid inflation is the interaction between the two fields. For simplicity, we assume a standard dimension 4 coupling of the form
\beq
V_{int}(\psi,\phi)={1\over2}g^2\psi^2\,\vec\phi\cdot\vec\phi .
\eeq
This term allows the waterfall field to be stabilized at $\phi=0$ at early times during slow-roll inflation when $\psi$ is displaced away from zero, and then to become tachyonic once the timer field approaches the origin; this is illustrated in Figure \ref{fig:Potential_Diagram}.

\subsection{Approximations}

We assume that the constant $V_0$ is dominant during inflation, leading to an approximate de Sitter phase with constant Hubble parameter $H$. By assuming a flat FRW background, the scale factor is approximated as 
\beq
a=\exp(Ht).
\eeq

At early times, $\psi$ is displaced from its origin, so $\phi=0$. This means that we can approximate the $\psi$ dynamics by ignoring the back reaction from $\phi$. The fluctuations in $\psi$ establish nearly scale invariant fluctuations on large scales, which we shall return to in Section \ref{Constraints}.  However, for the present purposes it is enough to treat $\psi$ as a classical, homogeneous field $\psi(t)$. We make the approximation that we can neglect the higher order terms in the potential $V_\psi$ in the transition era, leading to the equation of motion
\beq
\ddot \psi + 3H \dot \psi +\mpsi^2  \psi =0.
\label{psieqn}\eeq
Solving this equation for $\psi(t)$, we insert this into the equation for $\phi_i$.
We allow spatial dependence in $\phi_i$, and ignore, for simplicity, the higher order terms in $V_\phi$, leading to the equation of motion
\beq
\ddot \phi_i + 3H\dot \phi_i -e^{-2Ht} \nabla^2 \phi_i +m_\phi ^2 (t) \phi_i=0.
\label{phieqn}\eeq
Here we have identified an effective mass-squared for the waterfall field of
\beq
m_\phi^2(t) \equiv -\mphi^2 \left(1- \left ({ \psi(t) \over \psi_c} \right )^{\!2} \right),
\eeq
where the dimension 4 coupling $g$ has been traded for the parameter $\psi_c$ as $g^2=\mphi^2/\psi_c^2$. The quantity $\psi_c$ has the physical interpretation as the ``critical" value of $\psi$ such that the effective mass of the waterfall field passes through zero. So at early times for $\psi>\psi_c$, then $m_\phi^2>0$ and $\phi_i$ is trapped at $\phi_i=0$, while at late times for $\psi<\psi_c$, then $m_\phi^2<0$ and $\phi_i$ is tachyonic and can grow in amplitude, depending on the mode of interest; Figure \ref{fig:Potential_Diagram} illustrates these features.

\subsection{Mode Functions}

Since we ignore the back-reaction of $\phi$ onto $\psi$ and since we treat $\psi$ as homogeneous in the equation of motion for $\phi$ (eq.~(\ref{phieqn})), then by passing to $k$-space, all modes are decoupled.
Each waterfall field $\phi_i$ can be quantized and expanded in modes in momentum space as follows
\beq
\phi_i(\vec x,t) =
\int\! {d^3 k\over(2\pi)^3}  \left[c_{k,i} e^{i\kv \cdot \xv} u_k(t)+c_{k,i}^\dag e^{-i\kv \cdot \xv} u^*_k(t)\right]\!,
\label{eqn:phi(x)}
\eeq
where $c_k^\dagger\,(c_k)$ are the creation (annihilation) operators, acting on the $\phi_i=0$ vacuum.
By assuming an initial Bunch-Davies vacuum for each $\phi_i$, the mode functions $u_k$ are the same for all components due to the $SO(\N)$ symmetry. 
To be precise, we assume that at asymptotically early times the mode functions are the ordinary Minkowski space mode functions, with the caveat that we need to insert factors of the scale factor $a$ to convert from physical wavenumbers to comoving wavenumbers, i.e.,
\beq
u_k(t)\to{e^{-i\,k\,t/a}\over a\sqrt{2k}},\,\,\,\,\,\,\,\mbox{at early times}.
\eeq
At late times the mode functions behave very differently. Since the field becomes tachyonic, the mode functions grow exponentially at late times. The transition depends on the wavenumber of interest. The full details of the mode functions were explained very carefully in Ref.~\cite{sfak}, and the interested reader is directed to that paper for more information.

\begin{figure}[t]
\includegraphics[width=\cw]{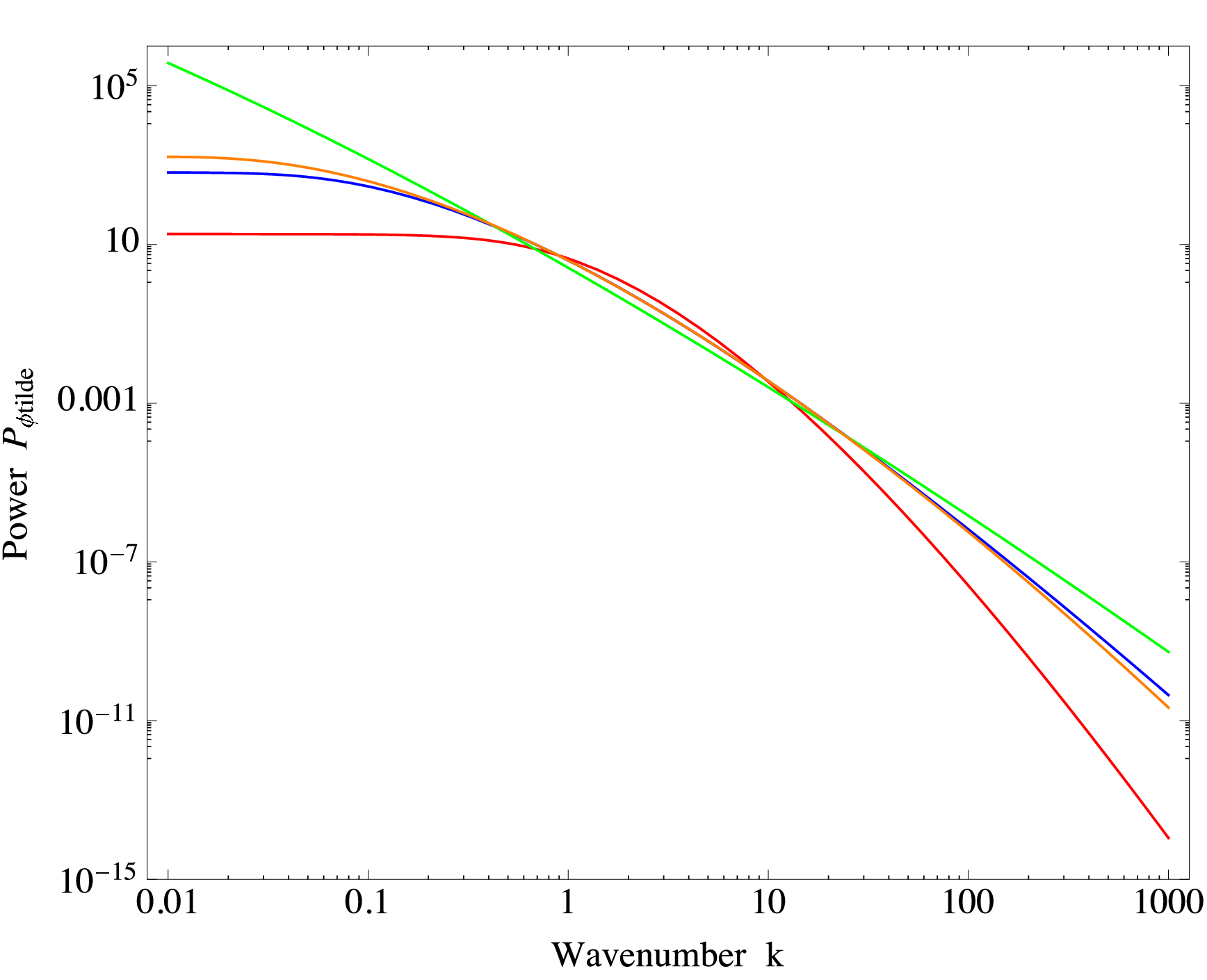}
\caption{A plot of the (re-scaled) field's power spectrum $P_{\tilde\phi}$ as a function of wavenumber $k$ (in units of $H$) for different masses: blue is $\mphi=2$ and $\mpsi=1/2$, red is $\mphi=4$ and $\mpsi=1/2$, green is $\mphi=2$ and $\mpsi=1/4$, orange is $\mphi=4$ and $\mpsi=1/4$.}
\label{fig:Pphi}
\end{figure}

Since we are approximating $\phi_i$ as a free field here, its fluctuations are entirely Gaussian and characterized entirely by its equal time two-point correlation function $\la \phi_i(\xv)\,\phi_j(\yv)\ra$. Passing to $k$-space, and using statistical isotropy and homogeneity of the Bunch-Davies vacuum, the fluctuations are equally well characterized by the so-called power spectrum $P_\phi(k)$, defined through 
\beq
\langle \phi_i(\kv_1)\,\phi_j(\kv_2)\rangle=(2\pi)^3\delta(\kv_1+\kv_2)\,\delta_{ij}\,P_\phi(k_1).
\eeq
This means the power spectrum is
\beq
\delta_{ij}\,P_\phi(k)=\int d^3x \, e^{-i{\bf k}\cdot{\bf x}}\langle \phi_i(\xv)\phi_j(\ov)\rangle.
\eeq
It is simple to show that the power spectrum is related to the mode functions by
\beq
P_\phi(k)=|u_k|^2.
\eeq
In Figure \ref{fig:Pphi} we plot a rescaled version of $P_\phi$, where we divide out by the root-mean-square (rms) of $\phi$ defined as $\phi_{rms}^2=\la\phi_i^2(0)\ra$. As will be mentioned in the next Section and is extensively discussed in \cite{sfak}, the ratio 
\beq
\tilde \phi (t) ={ \phi(t) \over \phi_{rms}(t)}
\eeq
is time-independent for late times, hence so is $P_{\tilde\phi}(k)$.

\section{The Time-Delay}\label{timedelay}

We would now like to relate the fluctuations in the waterfall field $\phi$ to a fluctuation in a physical observable, namely the density perturbation. An important step in this direction is to compute the so-called ``time-delay" $\dt(\xv)$; the time offset for the end of inflation for different parts of the universe. This causes different regions of the universe to have inflated more than others, creating a difference in their densities (though we will not explicitly compute $\delta\rho/\rho$ here). This basic formalism was first introduced by Hawking \cite{hawking} and by Guth and Pi \cite{guthpi1}, and has recently been reviewed in Ref.~\cite{guthsolvay}, where the transition from the time-delay formalism to the more frequently used curvature perturbation $\mathcal {R}$ is outlined. In the context of hybrid inflation, it was recently used by one of us in Ref.~\cite{sfak}.
It provides an intuitive and straightforward way to calculate primordial perturbations and we now use this to study perturbations established by the $\N$ waterfall fields. 

In its original formulation, the time-delay formalism starts by considering a classical homogeneous trajectory $\phi_0=\phi_0(t)$, and then considers a first order perturbation around this. At first order, one is able to prove that the fluctuating inhomogeneous field $\phi(\xv,t)$ is related to the classical field $\phi_0$, up to an overall time offset $\dt(\xv)$,
\beq
\phi(\xv, t ) =\phi_0( t -\dt(\xv)). \label{eq:tdelay}
\eeq

In the present case of hybrid inflation, the waterfall field is initially trapped at $\phi=0$ and then once it becomes tachyonic, it eventually falls off the hill-top due to quantum fluctuations. This means that there is no classical trajectory about which to expand. Nevertheless, we will show that, to a good approximation, the field $\phi(\xv,t)$ is well described by an equation of the form (\ref{eq:tdelay}), for a suitably defined $\phi_0$. The key is that all modes of interest grow at the same rate at late times. Further information of the time evolution of the mode functions can be found in Ref.~\cite{sfak}.

To show this, we need to compute the evolution of the field $\phi$ according to eq.~(\ref{phieqn}). This requires knowing $m_\phi(t)$, which in turn requires knowing $\psi(t)$. By solving eq.~(\ref{psieqn}) for the timer field and dispensing with transient behavior, we have
\beq
\psi(t) = \psi_c \, \exp\left(-\,p\,t\right),
\eeq
where
\beq
p = H\left({3\over2}-\sqrt{{9\over4}-{\mpsi^2\over H^2}}\right),
\label{pdef}\eeq
(note $p>0$). We have set the origin of time $t=0$ to be when $\psi=\psi_c$ and assume $\psi>\psi_c$ at early times.

Substituting this solution into $m_\phi(t)$, we can, in principle, solve eq.~(\ref{phieqn}). In general, the solution is somewhat complicated with a non-trivial dependence on wavenumber. However, at late times the behavior simplifies. Our modes of interest are super-horizon at late times. For these modes, the gradient term is negligible and the equation of motion reduces to
\beq
\ddot \phi_i + 3H\dot \phi_i +m_\phi ^2 (t) \phi_i=0.
\label{phieqn2}\eeq
So each mode evolves in the same way at late times. Treating $m_\phi(t)$ as slowly varying (which is justified because the timer field mass $\mpsi<H$ and so $p$ is small), we can solve for $\phi_i$ at late times $t$ in the adiabatic approximation. We obtain
\beq
\phi_i(\xv,t)=\phi_i(\xv,t_0)\exp\left(\int_{t_0}^t dt'\,\lambda(t')\right),
\label{phisoln}\eeq
where
\beq
\lambda(t) = H\left(-{3\over2}+\sqrt{{9\over4}+{m_0^2\over H^2}\big{(}1-e^{-2\,p\,t}\big{)}}\right).
\label{lambda}\eeq
Here $t_0$ is some reference time.
For $t>0$, we have $\lambda(t)>0$, so the modes grow exponentially in time. Later in Section \ref{Constraints} we explain that in fact $\lambda$ is roughly constant in the latter stage of the waterfall phase, i.e., the $\exp(-2\,p\,t)$ piece becomes small.

We now discuss fluctuations in the time at which inflation ends. For convenience, we define the reference time $t_0$ to be the time
at which the rms value of the field reaches $\phi_{end}$; the end of inflation
\beq
\N\phi_{\rm rms} ^2(t_0) = \phi^2_{\rm end},
\eeq
where we have included a factor of $\N$ to account for all fields, allowing $\phi_{rms}$ to refer to fluctuations in a single component $\phi_i$, i.e.,
$\phi_{rms}^2=\la\phi_i^2(0)\ra$. In terms of the power spectrum, it is 
\beq
\phi_{rms}^2=\int\!{d^3k\over(2\pi)^3}P_\phi(k).
\eeq
If we were to include arbitrarily high $k$, this would diverge quadratically, which is the usual Minkowski space divergence. However, our present analysis only applies to modes that are in the growing regime. For these modes, $P_\phi(k)$ falls off exponentially with $k$ and there is no problem. So in this integral, we only include modes that are in the asymptotic regime, or, roughly speaking, only super-horizon modes.

Using eq.~(\ref{phisoln}), we can express the field $\phi_i(\xv,t)$ at time $t = t_0 + \delta t$ in terms of the
field $\phi_i(\xv, t_0)$ by
\beq
\phi_i(\xv,t) = \phi_i(\xv,t_0) \exp\left(\lambda(t_0)\, \dt\right).
\eeq
If $t$ is chosen to be the time $t_{\rm end}(\xv)$ at which inflation ends at each point in space, then 
$\vec\phi\!\cdot\vec\phi\bigl(\xv,t_{\rm end}(\xv)\bigr) = \phi^2_{\rm end} = \N\phi_{\rm rms}^2(t_0)$, and the above equation becomes
\beq
\N\phi_{\rm rms}^2(t_0) = \vec\phi(\xv,t_0)\!\cdot\!\vec\phi(\xv,t_0) \exp\left(2\,\lambda(t_0) \, \dt\right)\!,
\eeq
which can be solved for the time-delay field $\dt(\xv) = t_{\rm end}(\xv) - t_0$ as
\beq
\dt(\xv) = {-1\over 2\lambda(t_0)} \ln\!\left ({ \vec\phi(\xv,t_0)\!\cdot\!\vec\phi(\xv,t_0) \over \N\phi_{\rm rms}^2(t_0) }\right ).
\label{dtfft}\eeq
This finalizes the $\N$ component analysis of the time-delay, generalizing the two component (complex) analysis of Ref.~\cite{sfak}.

\section{Correlation Functions}\label{correlation}

We now derive expressions for the two-point and three-point correlation functions of the time-delay field.  To do so, it is convenient to introduce a rescaled version of the correlation function $\Delta(x)$ defined through 
\beq
\la\phi_i(\xv) \phi_j(\ov)\ra = \Delta(x) \phi_{rms}^2 \delta_{ij}.
\eeq
By definition $\Delta(0)=1$, and as we vary $x$, $\Delta$ covers the interval $\Delta(x)\in(0,1]$.
(Ref.~\cite{sfak} used a different convention where $\Delta$ covers the interval $\Delta(x)\in(0,1/2]$).

\begin{figure}[t]
\includegraphics[width=\cw]{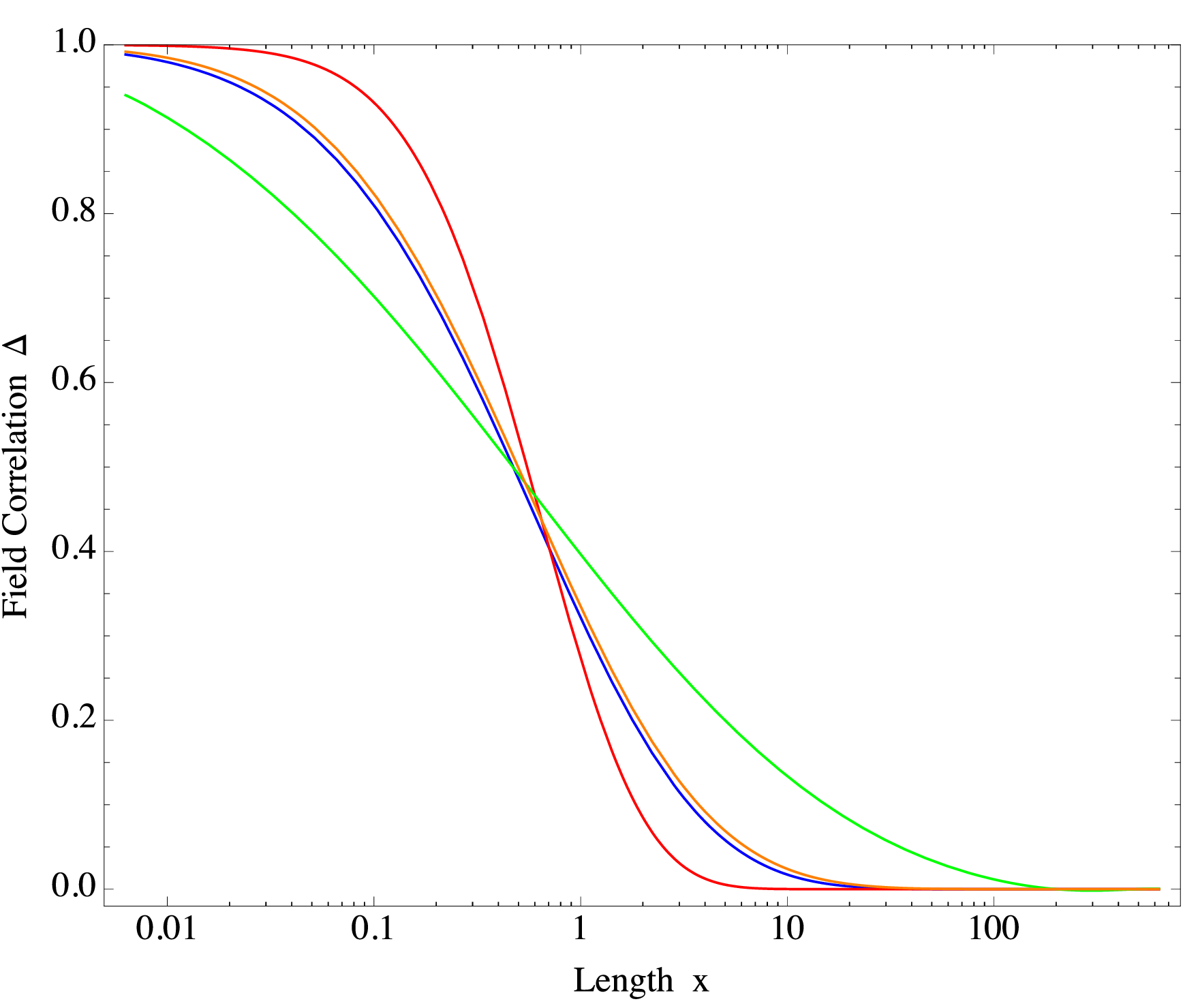}
\caption{A plot of the field's correlation $\Delta$ as a function of $x$ (in units of $H^{-1}$) for different masses: blue is $\mphi=2$ and $\mpsi=1/2$, red is $\mphi=4$ and $\mpsi=1/2$, green is $\mphi=2$ and $\mpsi=1/4$, orange is $\mphi=4$ and $\mpsi=1/4$.}
\label{fig:Delta_Vary}
\end{figure}

In Figure \ref{fig:Delta_Vary}, we show a plot of $\Delta$ at $t_{end}$ as a function of $x,$ measured in Hubble lengths ($H^{-1}$), for various combinations of masses.

\subsection{Two-Point Function}

We now express the time-delay correlation functions as a power series in $\Delta$ and $1/\N$. 
An alternative derivation of the power spectra of the time-delay field in terms of an integral, which is closer to the language of Ref.~\cite{sfak}, can be found in Appendix \ref{Integral}.

Using the above approximation for $\dt$ in eq.~(\ref{dtfft}), the two-point correlation function for the time-delay is
\bea
(2\lambda)^2\langle\dt(\xv)\dt(\ov)\rangle = \left\langle\!\ln\!\left({\vec \phi}_x\!\cdot\!{\vec \phi}_x\over \N\phi_{rms}^2\right)\ln\!\left({\vec \phi}_0\!\cdot\!{\vec \phi}_0\over \N\phi_{rms}^2\right)\!\right\rangle\!,\,\,\,\,\,\,\,\,
\label{deltatau}\eea
where we have used the abbreviated notation $\vec\phi_x\equiv\vec\phi(\xv)$.
The two-point function will include a constant (independent of $x$) for a non-zero $\la\dt\ra$. This can be reabsorbed into a shift in $t_0$, whose dependence we have suppressed here, and so we will ignore the constant in the following computation. This means that we will compute the {\em connected} part of the correlation functions.

We would like to form an expansion, but we do not have a classical trajectory about which to expand. Instead we use the following idea:
we recognize that ${\vec \phi}\cdot{\vec \phi}$ should be centralized around its mean value of $\N\phi_{rms}^2$, plus relatively small fluctuations at large $\N$. This means that it is convenient to write
\beq
{{\vec \phi}\cdot{\vec \phi}\over \N\phi_{rms}^2}=1+\left({{\vec \phi}\cdot{\vec \phi}\over \N\phi_{rms}^2}-1\right)
\eeq
and treat the term in parenthesis on the right as small, as it represents the fluctuations from the mean. This allows us to Taylor expand the logarithm in powers of 
\beq
\phii\equiv{{\vec \phi}\cdot{\vec \phi}\over \N\phi_{rms}^2}-1,
\eeq
with $\langle\phii\rangle=0$.
Now recall that the series expansion of the logarithm for small $\phii$ is
$\ln(1+\phii)=\phii-{\phii^2\over2}+{\phii^3\over3}-\ldots$,
allowing us to expand eq.~(\ref{deltatau}) to any desired order in $\phii$.

The leading non-zero order is quadratic $\sim\phii^2$. It is
\bea
(2\lambda)^2\langle\dt(\xv)\dt(\ov)\rangle_2\amp=\amp\langle\phii(\xv)\phii(\ov)\rangle,\nonumber\\
\amp=\amp{ \langle\phi_i(\xv)\phi_i(\xv)\phi_j(\ov)\phi_j(\ov)\rangle\over(\N\phi_{rms}^2)^2}-1,\,\,\,\,\,\,\,\,\,\,\,
\eea
where we are implicitly summing over indices in the second line (for simplicity, we will place all component indices ($i,\,j$) downstairs).
Using Wick's theorem to perform the four-point contraction, we find the result
\beq
(2\lambda)^2\langle\dt(\xv)\dt(\ov)\rangle_2={2\Delta^2(x)\over \N},
\label{2pt}
\eeq
where $x\equiv|\xv|$.
This provides the leading approximation for large $\N$, or small $\Delta$. This should be contrasted to the RSG approximation used in Refs.~\cite{sol1,sol2} and summarized in \cite{sfak}, in which the correlation function is approximated as $\sim\Delta$, rather than $\sim\Delta^2$, at leading order.

For brevity, we shall not go through the result at each higher order here, but we report on results at higher order in Appendix \ref{HigherOrder}.
By summing those results to different orders, we find
\bea
(2\lambda)^2\langle\dt(\xv)\dt(\ov)\rangle\amp=\amp{2\Delta^2\over \N}+{2\Delta^4\over \N^2}-{4\Delta^4\over \N^3}+{16\Delta^6\over 3 \N^3}+{8\Delta^4\over \N^4}\,\,\nonumber\\
\amp\amp-{32\Delta^6\over \N^4}+{24\Delta^8\over \N^4}+\mathcal{O}\!\left(1\over\N^{5}\right)\!, \,\,\,\,\,\,\,\,
\eea
(where $\Delta=\Delta(x)$ here).
We find that various cancellations have occurred. For example, the $-8\Delta^2/\N^2$ term that enters at cubic order (see Appendix \ref{HigherOrder}) has canceled.

Note that at a given order in $1/\N$ there are various powers of $\Delta^2$. However, by collecting all terms at a given power in $\Delta^2$, we can identify a pattern in the value of its coefficients as functions of $\N$. We find
\bea
\amp\amp\!\!\!\!(2\lambda)^2\langle\dt(\xv)\dt(\ov)\rangle \nonumber\\
\amp\amp= {2\Delta^2(x)\over \N}+{2\Delta^4(x)\over \N(\N+2)}+{16\Delta^6(x)\over 3\N(\N+2)(\N+4)}
\nonumber\\
\amp\amp\,\,\,\,\,\,+{24\Delta^8(x)\over \N(\N+2)(\N+4)(\N+6)}+\ldots
\label{FullExpansionTerms}\eea
We then identify the entire series as
\beq
(2\lambda)^2\langle\dt(\xv)\dt(\ov)\rangle=\sum_{n=1}^\infty C_n(\N)\,\Delta^{2n}(x),
\label{FullExpansion}
\eeq
where the coefficients are
\beq
C_n(\N)={1\over n^2}{{\N\over2}-1+n\choose n}^{\!-1},
\label{Coeff}\eeq
with ${a\choose b}$ the binomial coefficient. This series is convergent for any $\N$ and $\Delta\in(0,1]$, and is one of our central results.

For the case of a single scalar or a doublet (complex field), the full series organizes itself into known functions. 
For $\N=1$ we find
\bea
(2\lambda)^2\langle\dt(\xv)\dt(\ov)\rangle \amp =\amp 2\Delta^2(x)+{2\Delta^4(x)\over3}+{16\Delta^6(x)\over45}+\ldots\nonumber \\
\amp=\amp 2\,(\sin^{-1}\!\Delta(x))^2.
\eea
For $\N=2$ we find
\bea
(2\lambda)^2\langle\dt(\xv)\dt(\ov)\rangle \amp = \amp \Delta^2(x)+{\Delta^4(x)\over2^2}+{\Delta^6(x)\over3^2}+\ldots\nonumber \\
\amp = \amp \mbox{Li}_2(\Delta^2(x)),
\eea
where $\mbox{Li}_s(z)$ is the polylogarithm function.
We also find that for any even value of $\N$, the series is given by the polylogarithm function plus a polynomial in $\Delta$; this is described in Appendix \ref{EvenN}.

\begin{figure}[t]
\includegraphics[width=\cw]{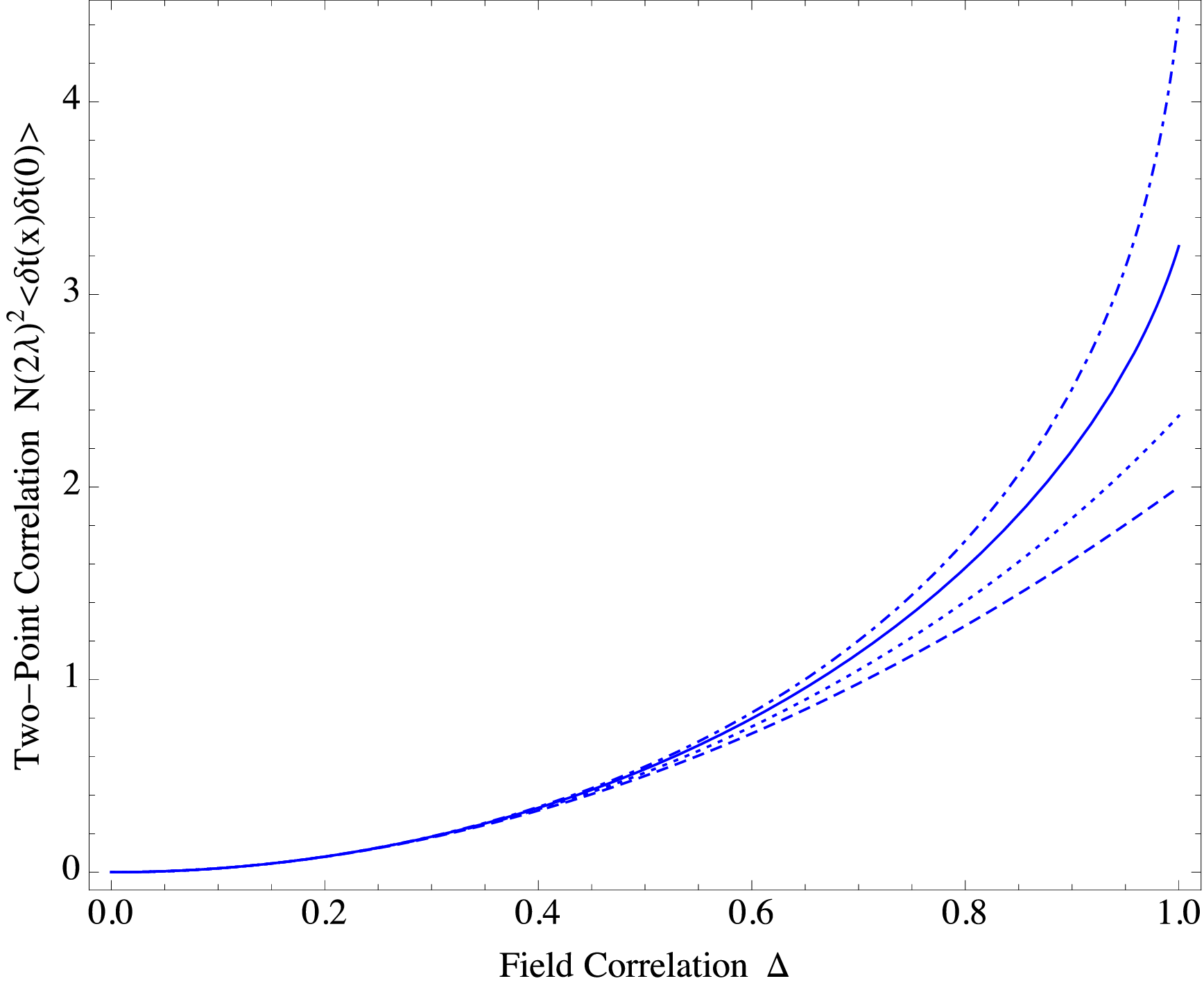}
\includegraphics[width=\cw]{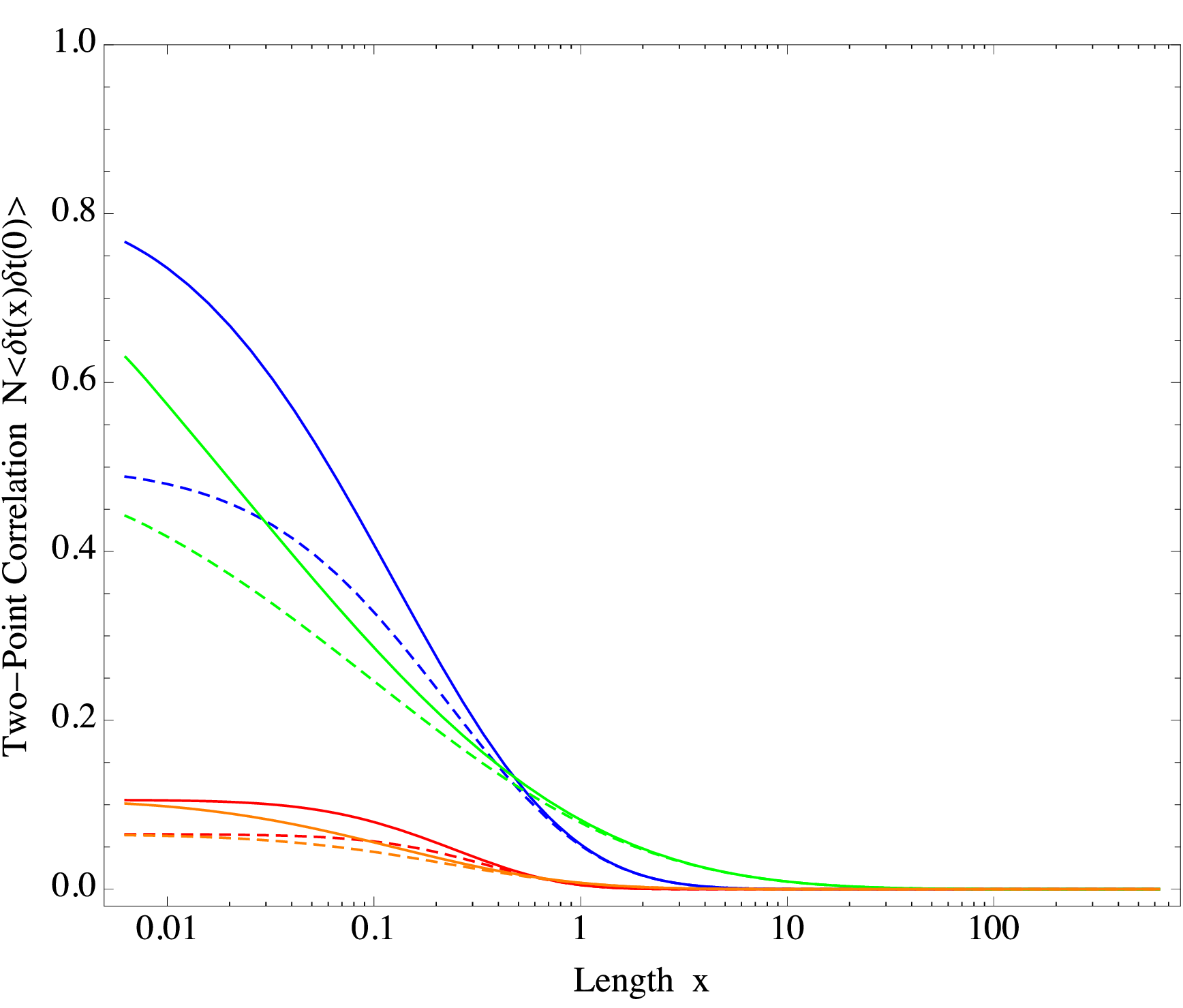}
\caption{Top: a plot of the (re-scaled) two-point function of the time-delay $\N(2\lambda)^2 \la\dt(\xv) \dt(\ov)\ra$ as a function of $\Delta\in[0,1]$ as we vary $\N$: dot-dashed is $\N=1$, solid is $\N=2$, dotted is $\N=6$, and dashed is $\N\to\infty$.
Bottom: a plot of the (re-scaled) two-point function of the time-delay $\N \la\dt(\xv) \dt(\ov)\ra$ as a function of $x$ for different masses: blue is $\mphi=2$ and $\mpsi=1/2$, red is $\mphi=4$ and $\mpsi=1/2$, green is $\mphi=2$ and $\mpsi=1/4$, orange is $\mphi=4$ and $\mpsi=1/4$, with solid for $\N=2$ and dashed for $\N\to\infty$.}
\label{fig:2pt}
\end{figure}

Using the power series, we can easily obtain plots of the two-point function for any $\N$. For convenience, we plot the re-scaled quantity $\N(2\lambda)^2 \la\dt(\xv) \dt(\ov)\ra$  as a function of $\Delta$ in Figure \ref{fig:2pt} (top) for different $\N$. We see convergence of all curves as we increase $\N$, which confirms that the leading behavior of the (un-scaled) two-point function is $\sim1/\N$. In Figure \ref{fig:2pt} (bottom) we plot $\N \la\dt(\xv) \dt(\ov)\ra$ as a function of $x$ for different masses and two different $\N$.

\subsection{Three-Point Function}

The three-point function is given by a simple modification of eq.~(\ref{deltatau}), namely
\bea
\amp\amp\!\!\!\!\!\!\!\!\!\!-(2\lambda)^3\langle\dt(\xv)\dt(\yv)\dt(\ov)\rangle \nonumber\\
\amp=\amp\left\langle\!\ln\!\left({\vec \phi}_x\!\cdot\!{\vec \phi}_x\over \N\phi_{rms}^2\right)
\ln\!\left({\vec \phi}_y\!\cdot\!{\vec \phi}_y\over \N\phi_{rms}^2\right)\ln\!\left({\vec \phi}_0\!\cdot\!{\vec \phi}_0\over \N\phi_{rms}^2\right)\!\right\rangle.\,\,\,\,\,\,\,\,\,
\label{deltatau3}\eea
Here we will work only to leading non-zero order, which in this case is cubic. We expand the logarithms as before to obtain
\bea
\amp\amp\!\!\!\!\!\!\!\!\!\!-(2\lambda)^3\langle\dt(\xv)\dt(\yv)\dt(\ov)\rangle_3 \nonumber\\
\amp=\amp\langle\phii(\xv)\phii(\yv)\phii(\ov)\rangle,\nonumber\\
\amp=\amp{\langle\phi_i(\xv)\phi_i(\xv)\phi_j(\yv)\phi_j(\yv)\phi_k(\ov)\phi_k(\ov)\rangle\over(\N\phi_{rms}^2)^3}+2\nonumber\\
\amp\amp-\left({\langle\phi_i(\xv)\phi_i(\xv)\phi_j(\yv)\phi_j(\yv)\rangle\over(\N\phi_{rms}^2)^2}+2\,\mbox{perms}\right),\,\,\,
\eea
where ``perms" is short for permutations under interchanging $\xv,\,\yv,\,\ov$.
Using Wick's theorem to perform the various contractions, we find the result
\beq
-(2\lambda)^3\langle\dt(\xv)\dt(\yv)\dt(\ov)\rangle_3 = {8\,\Delta(|\xv-\yv|)\,\Delta(x)\,\Delta(y)\over \N^2}.
\label{3pt}\eeq

We would now like to use the three-point function as a measure of non-Gaussianity.
For a single random variable, a measure of non-Gaussianity is to compute a dimensionless ratio of the skewness to the 3/2 power of the variance. For a field theory, we symmetrize over variables, and define the following measure of non-Gaussianity in position space
\beq
S\equiv{\langle\dt(\xv)\dt(\yv)\dt(\ov)\rangle\over\sqrt{\langle\dt(\xv)\dt(\yv)\rangle\langle\dt(\xv)\dt(\ov)\rangle\langle\dt(\yv)\dt(\ov)\rangle}}.
\label{NGdef}\eeq
(Recall that we are ignoring $\la\dt\ra$, as it can be just re-absorbed into $t_0$, so the three-point and two-point functions are the connected pieces).
Computing this at the leading order approximation using eqs.~(\ref{2pt},\,\ref{3pt}) (valid for large $\N$, or small $\Delta$), we find
\beq
S\approx-\sqrt{8\over \N}\,.
\eeq
Curiously, the dependence on $\xv,\,\yv$ has dropped out at this order. We see that for small $\N$ there is significant non-Gaussianity, while for large $\N$ the theory becomes Gaussian, as expected from the central limit theorem.

\subsection{Momentum Space}

Let us now present our results in $k$-space. We shall continue to analyze the results at high $\N$, or small $\Delta$, which allows us to just include the leading order results.

\begin{figure}[t]
\includegraphics[width=\cw]{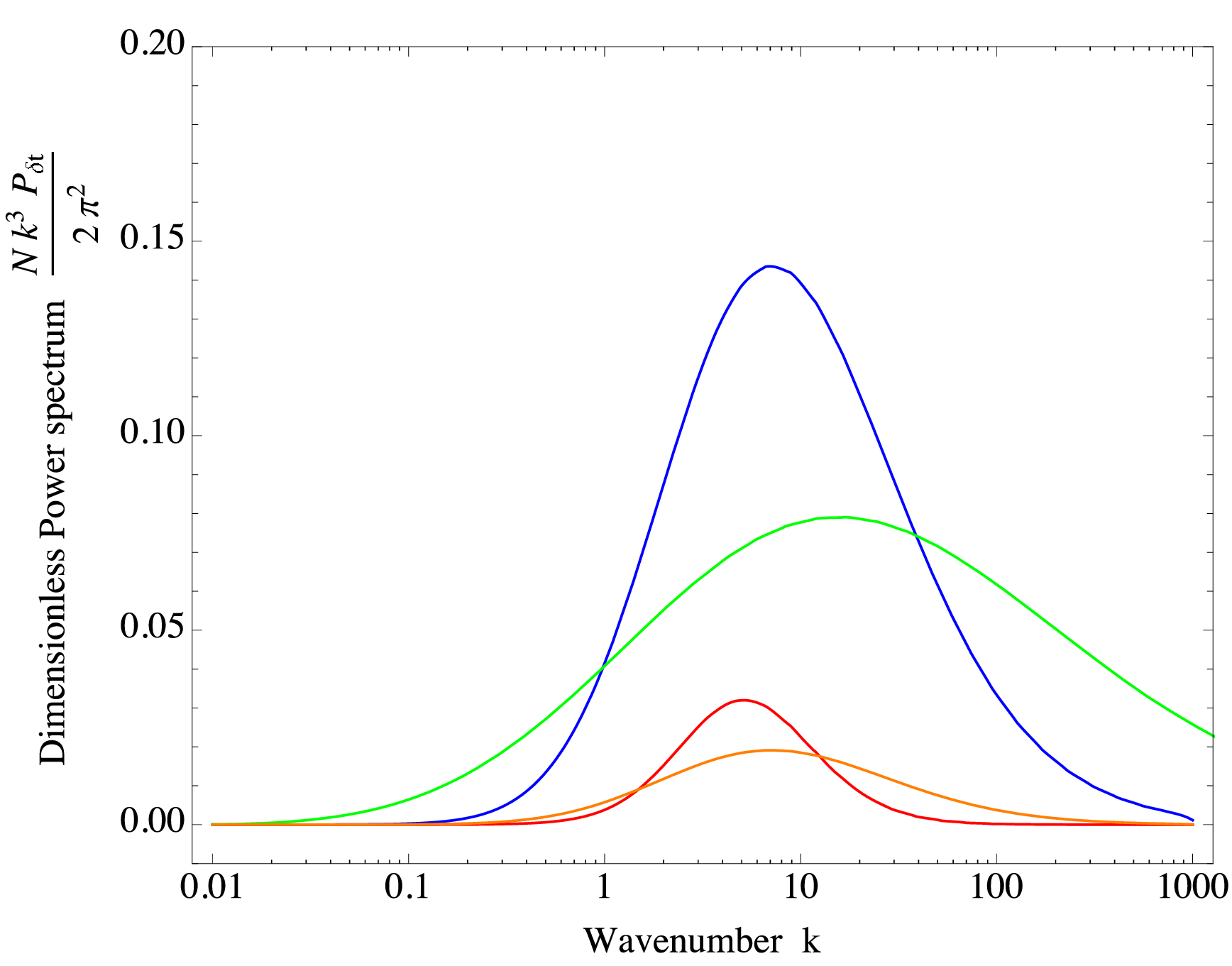}
\caption{The dimensionless power spectrum $\N\mathcal{P}_{\dt}(k)$ at large $\N$ as a function of $k$ (in units of $H$) for different choices of masses: blue is $\mphi=2$ and $\mpsi=1/2$, red is $\mphi=4$ and $\mpsi=1/2$,  is $\mphi=2$ and $\mpsi=1/4$, orange is $\mphi=4$ and $\mpsi=1/4$.}
\label{fig:PowerSpectrum}
\end{figure}

For the two-point function, we define the power spectrum through
\beq
\langle \dt(\kv_1)\dt(\kv_2)\rangle=(2\pi)^3\delta(\kv_1+\kv_2)P_{\dt}(k_1).
\eeq
We use eq.~(\ref{2pt}) and Fourier transform to $k$-space using the convolution theorem. 
To do so it is convenient to introduce a dimensionless field $\tilde{\phi}_i\equiv\phi_i/\phi_{rms}$ and introduce the corresponding power spectrum $P_{\tilde{\phi}}(k)=P_{\phi}(k)/\phi_{rms}^2=|u_k|^2/\phi_{rms}^2$, which is the Fourier transform of $\Delta(x)$.
We then find the result
\beq
P_{\dt}(k)\approx{1\over 2\lambda^2\N}\int\!{d^3k'\over(2\pi)^3}P_{\bar{\phi}}(k')P_{\bar{\phi}}(|\kv-\kv'|).
\label{Ptau}\eeq
A dimensionless measure of fluctuations is the following
\beq
\mathcal{P}_{\dt}(k)\equiv {k^3 H^2 P_{\dt}(k)\over2\pi^2},
\eeq
The factor of $k^3/(2\pi^2)$ is appropriate as this gives the variance per log interval: $\la(H\dt)^2\ra=\int\!d\ln k\,\,\mathcal{P}_{\dt}(k)$.
By studying eq.~(\ref{Ptau}), one can show $P_{\dt}\approx$\,const for small $k$ and falls off for large $k$. This creates a spike in $\mathcal{P}_{\dt}(k)$ at a finite $k^*$ and its amplitude is rather large. (This is to be contrasted with the usual fluctuations in de Sitter space, $P_{\dt}(k)\propto 1/k^3$, making $\mathcal{P}_{\dt}(k)$ approximately scale invariant.)
We see that the amplitude of the spike scales as $\sim 1/\N$, and so it is reduced for large $\N$. A plot of $\mathcal{P}_{\dt}(k)$ is given in Figure \ref{fig:PowerSpectrum}.

For the three-point function, we define the bispectrum through
\beq
\langle \dt(\kv_1)\dt(\kv_2)\dt(\kv_3)\rangle=(2\pi)^3\delta(\kv_1+\kv_2+\kv_3)B_{\dt}(k_1,k_2,k_3)
\eeq
where we have indicated that the bispectrum only depends on the magnitude of the 3 $k$-vectors, with the constraint that the vectors sum to zero.
We use eq.~(\ref{3pt}) and Fourier transform to $k$-space, again using the convolution theorem. We find
\bea
\amp\amp \!\!\! B_{\dt}(k_1,k_2,k_3) \approx  -{1\over3\lambda^3\N^2}\times \nonumber\\
\amp\amp \!\left[\int\!{d^3k\over(2\pi)^3}P_{\bar{\phi}}(k)P_{\bar{\phi}}(|\kv_1-\kv|)P_{\bar{\phi}}(|\kv_2+\kv|)+2\,\mbox{perms}\right]\,\,\,\,\,\,\,\,\,\,
\label{Btau}\eea

\begin{figure}[t]
\includegraphics[width=\cw]{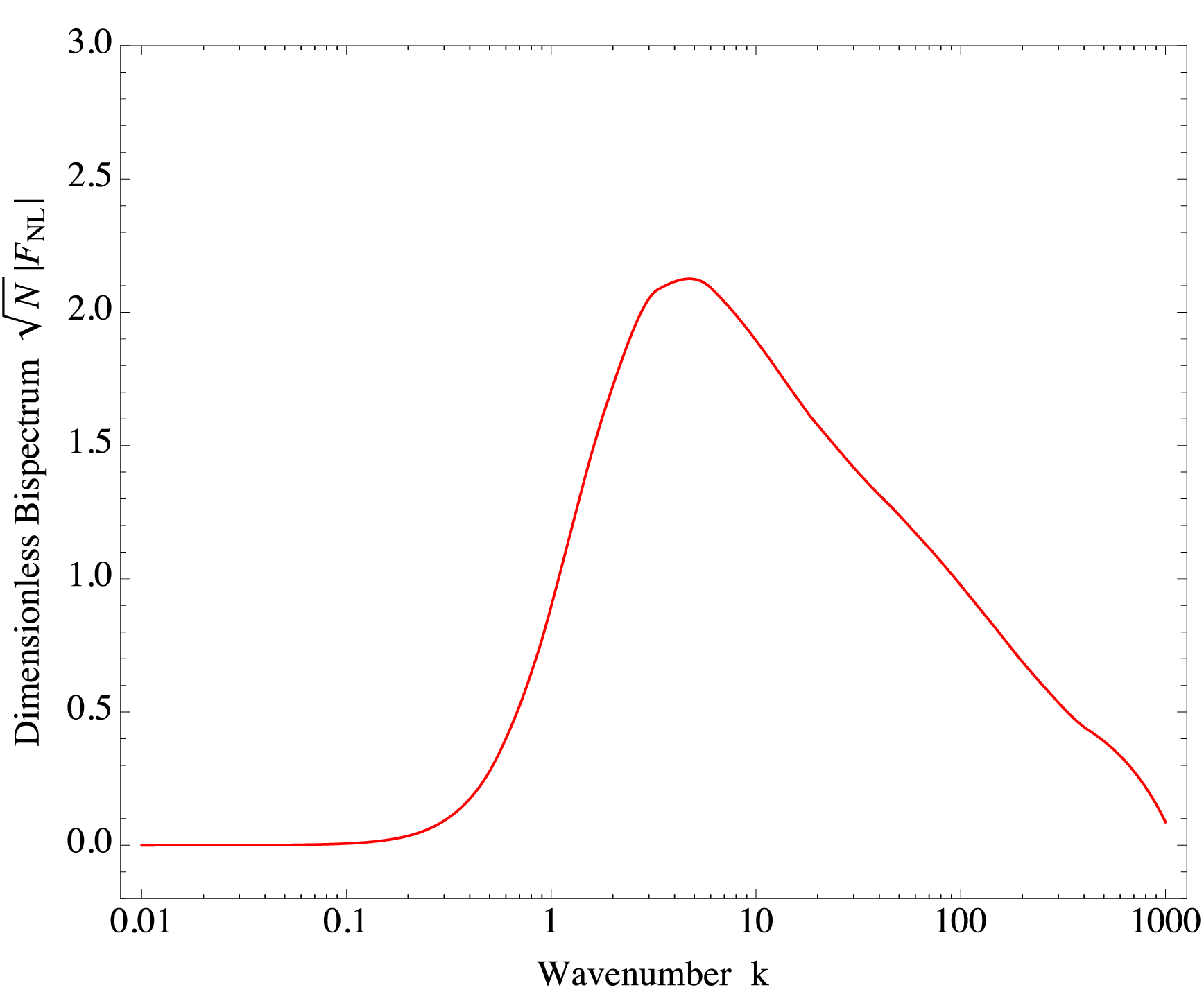}
\caption{The dimensionless bispectrum $\sqrt{\N}\,\mathcal{F}_{NL}$ as a function of $k$ (in units of $H$) at large $\N$ for $\mphi=4$ and $\mpsi=1/2$.}
\label{fig:Bispectrum}
\end{figure}

To measure non-Gaussianity in $k$-space, it is conventional to introduce the dimensionless $f_{NL}$ parameter, defined as\footnote{A factor of 6/5 is often included when studying the gauge invariant quantity $\zeta$ that appears in cosmological perturbation theory, but it does not concern us here.}
\beq
f_{NL}(k_1,k_2,k_3)\equiv{B_{\dt}(k_1,k_2,k_3)\over P_{\dt}(k_1)P_{\dt}(k_2)+2\,\mbox{perms}}.
\eeq
By substituting the above expressions for $P_{\dt}$ and $B_{\dt}$, we see that  $f_{NL}$ is independent of $\N$ at this leading order.
However, this belies the true dependence of non-Gaussianity on the number of fields. This is because $f_{NL}$ is a quantity that can be large even if the non-Gaussianity is relatively small (for example, on CMB scales, any $f_{NL}$ smaller than $\mathcal{O}(10^{5})$ is a small level of non-Gaussianity).
Instead a more appropriate measure of non-Gaussianity in $k$-space is to compute some ratio of the bispectrum to the 3/2 power of the power spectrum, analogous to the position space definition in eq.~(\ref{NGdef}). For the simple equilateral case, $k_1=k_2=k_3$, we define
\beq
\mathcal{F}_{NL}(k) \equiv {k^{3/2} B_{\dt}(k)\over 3\sqrt{2}\,\pi P_{\dt}(k)^{3/2}},
\eeq
where we have inserted a factor of $k^{3/2}\!/(\sqrt{2}\,\pi)$ from measuring the fluctuations per log interval.
Using eqs.~(\ref{Ptau},\,\ref{Btau}), we see that $\mathcal{F}_{NL}\propto 1/\sqrt{\N}$, as we found in position space.
In Figure \ref{fig:Bispectrum}, we plot this function. We note that although the non-Gaussianity can be large, the peak is on a length scale that is small compared to the CMB and so it evades recent bounds \cite{Ade:2013uln}.

\section{Constraints on Hybrid Models}\label{Constraints}

Hybrid inflation models must satisfy several observational constraints. Here we discuss these constraints, including the role that $\N$ plays, and discuss the implications for the scale of the density spike.

\subsection{Topological Defects}

The first constraint on hybrid models concerns the possible formation of topological defects. Since the waterfall field starts at $\phi=0$ and then falls to some vacuum, it spontaneously breaks a symmetry. For a single field $\N=1$, this breaks a $\mathbb{Z}_2$ symmetry; see Figure \ref{fig:Potential_Diagram}. For multiple fields $\N>1$, this breaks an $SO(\N)$ symmetry. 
$\N=1$ leads to the formation of {\em domain walls}, which are clearly ruled out observationally, so these models are strongly disfavored; 
$\N=2$  leads to the formation of {\em cosmic strings}, which have not been observed and if they exist are constrained to be small in number. The subject of cosmic string production in hybrid inflation and their subsequent effects is discussed in \cite{lindeBH}. For current bounds on cosmological defects from the Planck collaboration the reader can refer to \cite{Ade:2013xla} and references therein. Diluting cosmic strings to make them unobservable in our case would require a very large number of e-foldings of the waterfall phase to make compatible with observations, and seems unrealistic. Further increasing the number of waterfall fields,
$\N=3$ leads to the formation of {\em monopoles}, which are somewhat less constrained;
$\N=4$ leads to the formation of {\em textures}, which are relatively harmless;
$\N>4$ avoids topological defects altogether.
So choosing several waterfall fields is preferred by current constraints on topological defects.

Even for $\N>4$, one might be concerned about constraints from the re-ordering of $\N-1$ Goldstone modes arising from the breaking of the $SO(\N)$ global symmetry. However it is important to note that all global symmetries are only ever approximate. So it is expected that these modes are not strictly massless, but pick up a small mass at some order, as all Goldstones do. The only important point is that the mass of the Goldstones $m_G$  is small compared to the mass scales of the inflaton and the waterfall field. For example, if we have a scale of inflation with $H\sim10^{12}$\,GeV, then the inflaton and waterfall fields should have a mass not far from $\sim10^{12}$\,GeV also. We then only need the Goldstone masses to satisfy $m_G\ll 10^{12}$\,GeV for our analysis to be true, and for the $SO(\N)$ symmetry to be a good approximation for the purpose of the phase transition. At late times, these (approximate) Goldstones will appear heavy and relax to the bottom of their potential, and be harmless. So, with this in mind, a large VEV for the waterfall field is indeed observationally allowed, and large e-folds after the transition leading to large black holes is indeed allowed. 

\subsection{Inflationary Perturbations}

Inflation generates fluctuations on large scales which are being increasingly constrained by data.
An important constraint on any inflation model is the bound on the tensor-to-scalar ratio $r$. CMB measurements from Planck place an upper bound on tensor modes of $r<0.11$ (95\% confidence) \cite{Ade:2013uln}. The amplitude of tensor modes is directly set by the energy density during inflation. Typical hybrid models are at relatively low energy scales, without the need for extreme fine tuning, and so they immediately satisfy this bound.
Recent data by the BICEP2 experiment \cite{bicep} claim a detection of gravitational waves with $r\sim 0.2$ at a high confidence level. Going back to the simplest potential of hybrid inflation $V(\psi) = V_0 + {1\over 2} m_\psi^2 \psi^2_...$, we have so far considered the regime $V_0\gg {1\over 2} m_\psi^2 \psi^2$, which generates a very small amount of tensor modes. However one can operate in a regime with $V_0 \gtrsim {1\over 2} m_\psi^2 \psi^2$, where the tensor to scalar ratio can be pushed to be closer to ${\cal O}(0.1)$. Furthermore, new realizations of hybrid inflation \cite{Hertzberg:2014sza,Carone:2014lba} can generate appreciable tensor modes with appropriate parameter choices. 
Since subtracting the dust foreground has reduced the detected signal to an updated upper bound on the tensor modes \cite{bicep_planck,Ade:2015tva}, 
we will leave a complete discussion of producing the correct value of $r$ in hybrid inflation models for future work.

Although the detection of tensor modes is not confirmed
, scalar modes are pinned down with great accuracy. The tilt of the scalar mode spectrum is characterized by the primordial spectral index $n_s$. WMAP \cite{Hinshaw:2012aka} and Planck measurements \cite{Ade:2013uln} place the spectral index near 
\beq 
n_{s,obs} \approx 0.96, 
\eeq 
giving a red spectrum. Here we examine the constraints imposed on hybrid models in order to obtain this value of $n_s$.

The tilt on large scales is determined by the timer field $\psi$. For low scale models of inflation, such as hybrid inflation, the prediction for the spectral index is
\beq
n_s=1+2\,\eta,
\eeq
where
\beq
\eta\approx{1\over8\pi G}{V_\psi''\over V_0}={V_\psi''\over3\,H^2}.
\eeq
This is to be evaluated $N_e$ e-foldings from the end of inflation, where $N_e=50-60$ in typical models. Combining the above equations, we need to satisfy $V_\psi''\approx-0.06\,H^2$. If we take $V_\psi=\mpsi^2\psi^2/2$, then $V_\psi''>0$, and $n_s>1$, which is ruled out. So we need higher order terms in the potential to cause it to become concave down at large values of $\psi$ where $\eta$ is evaluated, while leaving the quadratic approximation for $V_\psi$ valid at small $\psi$. For most reasonable potential functions, such as potentials that flatten at large field values, we expect $|V_\psi''|\lesssim\mpsi^2$. So this suggests a bound 
\beq
\mpsi^2\gtrsim0.06\, H^2,
\label{mpsibd}\eeq
which can only be avoided by significant fine tuning of the potential. Hence although the timer field is assumed light ($\mpsi<H$), it cannot be extremely light.

As an example let us consider a potential that can be approximated by $V(\psi) = {1\over 2} m_\psi ^2 \psi^2 -{g^2 \over 4} \psi^4$. We are neglecting higher order terms needed to stabilize the potential. Let us take $V''(\psi_{\rm CMB}) = m_\psi^2 - 3g^2 \psi_{\rm CMB}^2 <0$, where $\phi_{\rm CMB}$ is the field value at the point where the CMB fluctuations exit the horizon during inflaiton, in order to produce a red-tilted spectrum. This requirement leads to $m_\psi^2 \lesssim 3g^2 \psi_{\rm CMB}^2$ which in turns implies $V''(\psi_{\rm CMB}) \sim m_\psi^2$. Since fluctuations that are imprinted on the CMB must exit the horizon sufficiently before the waterfall transition, in order for the CMB spectrum to keep its approximate scale-invariance, the quartic term in the potential considered in this example will be subdominant during the waterfall transition, since $\psi_{\rm waterfall} < \psi_{\rm CMB}$. 
An example potential for the timer field is 
\beq
V_\psi(\psi)={F^2m_\psi^2\over2} \ln(1+\psi^2/F^2).
\eeq
 If one Taylor expands this, one finds the needed $m_\psi^2\psi^2/2$ quadratic piece, plus the higher order corrections that render the potential {\em concave down}, which is compatible with the observed red-tilted spectrum.
 
Since a detailed calculation of the primordial black hole production for specific realizations of our model (along with connecting it to potential observables like supermassive black holes) is deferred for a future presentation, the corresponding choice of potential parameters will be done at that time as well.

\subsection{Implications for Scale of Spike}

The length scale associated with the spike in the spectrum is set by the Hubble length during inflation $H^{-1}$, red-shifted by the number of e-foldings of the waterfall phase $\Nw$. Since the Hubble scale during inflation is typically microscopic, we need the duration of the waterfall phase $\Nw$ to be significant (e.g., $\Nw\sim 30-40$) to obtain a spike on astrophysically large scales. Here we examine if this is possible.

Since we have defined $t=0$ to be when the transition occurs ($\psi=\psi_c$), then $\Nw=H t$ with final value at $t=t_{end}$. 
To determine the final value, we note that modes grow at the rate $\lambda$, derived earlier in eq.~(\ref{lambda}). For $\mpsi<H$, we can approximate the parameter $p$ (eq.~(\ref{pdef})) as $p\approx\mpsi^2/3H$. Using the above spectral index bound in eq.~(\ref{mpsibd}), we see that for significantly large $\Nw$ (e.g., $\Nw\sim 30-40$) the exponential factor 
\beq
\exp(-2\,p\,t)\approx\exp\left(-{2\,\mpsi^2\Nw\over3\,H^2}\right) 
\eeq
is somewhat small and we will ignore it here. In this regime, the dimensionless growth rate $\lambda/H$ can be approximated as a constant
\beq
{\lambda\over H}\approx -{3\over2}+\sqrt{{9\over4}+{\mphi^2\over H^2}}.
\label{lambdaapprox}\eeq
The typical starting value for $\phi$ is roughly of order $H$ (de Sitter fluctuations) and the typical end value for $\phi$ is roughly of order $\mpl$ (Planck scale). For self consistency, $\phi$ must pass from its starting value to its end value in $\Nw$ e-foldings with rate set by $\lambda/H$. This gives the approximate value for $\Nw$ as\footnote{A better approximation comes from tracking the full time dependence of $\lambda$ and integrating the argument of the exponential exp($\int^t dt'\lambda(t'))$, but this approximation suffices for the present discussion.}
\beq
\Nw\approx {H\over\lambda}\ln\!\left(\mpl\over H\right).
\eeq
This has a clear consequence: If we choose $\mphi\gg H$, as is done in some models of hybrid inflation, then $H/\lambda\ll1$. 
So unless we push $H$ to be many, many orders of magnitude smaller then $\mpl$, then $\Nw$ will be rather small. This will lead to a spike in the spectrum on rather small scales and possibly ignorable to astrophysics.

Note that if we had ignored the spectral index bound that leads to eq.~(\ref{mpsibd}), then we could have taken $\mpsi$ arbitrarily small, leading to an arbitrarily small $p$ value. In this (unrealistic) limit, it is simple to show
\beq
{\lambda\over H}\approx {2\,\mpsi^2\,\mphi^2\,\Nw\over 9\,H^4}.
\eeq
So by taking $\mpsi$ arbitrarily small, $\lambda$ could be made small, and $\Nw$ could easily be made large. 
However, the existence of the spectral index bound essentially forbids this, requiring us to go in a different direction.

The only way to increase $\Nw$ and satisfy the spectral index bound on $\mpsi$ is to take $\mphi$ somewhat close to $H$. This allows $H/\lambda$ to be appreciable from eq.~(\ref{lambdaapprox}). For instance, if we set $\mphi=1.3\,H$, then $H/\lambda\approx 2$. If we then take $H$ just a few orders of magnitude below $\mpl$, say $H\sim10^{-6,7}\mpl$, which is reasonable for inflation models, we can achieve a significant value for $\Nw$. This will lead to a spike in the spectrum on astrophysically large scales, which is potentially quite interesting. It is possible that there will be distortions in the spectrum by taking $\mphi$ close to $H$, but we will not explore those details here. However, there is an important consequence that we explore in the next subsection.

\subsection{Eternal Inflation}

Since we are being pushed towards a somewhat low value of $\mphi$, near $H$, we need to check if the theory still makes sense. One potential problem is that the theory may enter a regime of eternal inflation. This could occur for the waterfall field at the hill-top. This would wipe out information of the timer field, which established the approximately scale invariant spectrum on cosmological scales.

The boundary for eternal inflation is roughly when the density fluctuations are $\mathcal{O}(1)$, and this occurs when the fluctuations in the time delay are $\la(H\,\dt)^2\ra=\mathcal{O}(1)$. To convert this into a lower bound on $\mphi$, let us imagine that $\mphi$ is even smaller than $H$. In this regime, the growth rate $\lambda$ can be estimated using eq.~(\ref{lambdaapprox}) as $\lambda\sim \mphi^2/H^2$. Using eq.~(\ref{2pt}) this gives $\la(H\,\dt)^2\ra\sim H^4/(\N\mphi^4)$. This implies that eternal inflation occurs when the waterfall mass is below a critical value $\mphi^c$, which is roughly
\beq
\mphi^c\sim{H\over\N^{1/4}}.
\eeq
So when $\N\sim1$ we cannot have $\mphi$ near $H$, because we then enter eternal inflation. On the other hand, for large $\N$ we are allowed to have $\mphi$ near $H$ and avoid this problem. This makes sense intuitively, because for many fields it is statistically favorable for at least one of the fields to fall off the hill-top, causing inflation to end. Hence large $\N$ is more easily compatible with the above set of constraints than low $\N$.

\section{Discussion and Conclusions}\label{discussion}

In this work we studied density perturbations in hybrid inflation caused by $\N$ waterfall fields, which contains a spike in the spectrum. 
We derived expressions for correlation functions of the time-delay and constrained parameters with observations.
 
{\em Density Perturbations}:
We derived a convergent series expansion in powers of $1/\N$ and $\Delta(x)$, the dimensionless correlation function for the field, for the two-point function of the time-delay for any $\N$, and the leading order behavior of the three-point function of the time-delay for large $\N$.
These correlation functions are well approximated by the first term in the series for large $\N$ (even for $\N=2$ the leading term is moderately accurate to $\sim30\%$, and much more accurate for higher $\N$). In this regime, the fluctuations are suppressed, with two-point and three-point functions given by
\bea
\la \dt(\xv)\,\dt(\ov)\ra \amp \approx \amp {\Delta^2(x)\over2\lambda^2\N},\nonumber\\
\la \dt(\xv)\,\dt(\yv)\,\dt(\ov)\ra \amp \approx \amp - {\Delta(|\xv-\yv|)\Delta(x)\Delta(y)\over\lambda^3\N^2}.
\eea
Although this reduces the spike in the spectrum, for any moderate value of $\N$, such as $\N=3,4,5,$ the amplitude of the spike is still quite large (orders of magnitude larger than the $\sim 10^{-5}$ level fluctuations on larger scales relevant to the CMB), and may have significant astrophysical consequences. Also, the relative size of the three-point function to the 3/2 power of the two-point function scales as $\sim1/\sqrt{\N}$. In accordance with the central limit theorem, the fluctuations become more Gaussian at large $\N$. This will make the analysis of the subsequent cosmological evolution more manageable, as this provides a simple spectrum for initial conditions. 
We note that since we are considering small scales compared to the CMB, then this non-Gaussianity evades Planck bounds \cite{Ade:2013uln}.

{\em Constraints}:
We mentioned that hybrid models avoid topological defects for large $\N$, while tensor mode constraints can be satisfied in different models.
A very serious constraint on hybrid models comes from the observed spectral index $n_s\approx 0.96$, which requires the potential to flatten at large field values. One consequence of this is that the timer field mass $\mpsi$ needs to be only a little smaller than the Hubble parameter during inflation $H$, or else the model is significantly fine tuned. 
For a large value of the waterfall field mass $\mphi$, this would imply a large growth rate of fluctuations, a rapid termination of inflation, and in turn a density spike on very small scales. 
Otherwise, we need to make the waterfall field mass $\mphi$ somewhat close to $H$, but this faces problems with eternal inflation. However, by using a large number of waterfall fields $\N$, it is safer to make the waterfall field mass $\mphi$ somewhat close to $H$. This reduces the growth rate of fluctuations, prolonging the waterfall phase for many e-foldings.

\begin{itemize}
\item[] Thus large $\N$ presents a plausible setup to establish a spike in the density perturbations on astrophysically large length scales that is consistent with other constraints.
\end{itemize}

{\em Outlook}:
It may be possible that these perturbations seed primordial black holes, which may be relevant to seeding supermassive black holes, or an intriguing form of dark matter. Since black hole formation is exponentially sensitive on the inflationary power spectrum, an accurate calculation of fluctuations is important, but predicting astrophysical observables does not follow trivially.
An investigation into these topics is underway. 
It would be important to fully explore the eternal inflation bound and the effects on the spectrum for relatively light waterfall field masses.
Finally, it would be of interest to try to embed these large $\N$ models into fundamental physics, such as string theory, and to explore reheating 
\cite{kofmanreheat,Greene:1997fu,Felder:2000hj,Hertzberg:2014iza} 
and baryogenesis \cite{Copeland:2001qw,Delepine:2006rn,baryon1,baryon2,Hertzberg:2014jza,Hook:2014mla,Lozanov:2014zfa} in this framework.

\vspace{0.3cm}
\begin{center}{\bf Acknowledgments}\end{center}
We would very much like to thank Alan Guth for his guidance and contribution to this project. 
We would also like to thank Victor Buza and Alexis Giguere for very helpful discussions. 
This work is supported by the U.S. Department of Energy under grant Contract Number  DE-SC00012567
and in part by MIT's Undergraduate Research Opportunities Program. EIS gratefully acknowledges support from a Fortner Fellowship at UIUC.
Research at Perimeter Institute is supported by the Government of Canada through Industry Canada and by the Province of Ontario through the Ministry of Research and Innovation.

\begin{appendix}

\section{Series Expansion to Higher Orders}\label{HigherOrder}

Earlier we computed the two-point correlation function for the time-delay at quadratic order, and then stating the results at all orders. Here we mention the results order by order.

\vspace{0.1cm}\begin{center}{\em a. Cubic Order}\end{center}\vspace{0.2cm}

At cubic order $\sim\phii^3$ we find 
\bea
(2\lambda)^2\langle\dt(\xv)\dt(\ov)\rangle_3
\amp=\amp-{8\Delta^2\over \N^2}
\eea

\vspace{0.1cm}\begin{center}{\em b. Quartic Order}\end{center}\vspace{0.2cm}

At quartic order $\sim\phii^4$ we find 
\bea
(2\lambda)^2\langle\dt(\xv)\dt(\ov)\rangle_4
\amp=\amp{8\Delta^2+2\Delta^4\over \N^2}+{40\Delta^2+4\Delta^4\over \N^3}\,\,\,
\eea

\vspace{0.1cm}\begin{center}{\em c. Quintic Order}\end{center}\vspace{0.2cm}

At quintic order $\sim\phii^5$ we find 
\bea
(2\lambda)^2\langle\dt(\xv)\dt(\ov)\rangle_5
\amp=\amp-{96\Delta^2+32\Delta^4\over \N^3}\nonumber\\
\amp\amp-{256\Delta^2+64\Delta^4\over \N^4}
\eea

\vspace{0.1cm}\begin{center}{\em d. Sextic Order}\end{center}\vspace{0.2cm}

At sextic order $\sim\phii^6$ we find 
\bea
(2\lambda)^2\langle\dt(\xv)\delta\tau(\ov)\rangle_6
\amp=\amp{168\Delta^2+72\Delta^4+16\Delta^6\over 3\N^3}\nonumber\\
\amp\amp+{1056\Delta^2+464\Delta^4+32\Delta^6\over \N^4}\nonumber\\
\amp\amp+\,\, {6144\Delta^2+2496\Delta^4+128\Delta^6\over3\N^5}\,\,\,\,\,\,\,\,\,\,\,
\eea

\vspace{0.1cm}\begin{center}{\em e. Septic Order}\end{center}\vspace{0.2cm}

At septic order $\sim\phii^7$ we find 
\bea
(2\lambda)^2\langle\dt(\xv)\dt(\ov)\rangle_7\amp=\amp
-{32(43\Delta^2+22\Delta^4+6\Delta^6)\over \N^4}\nonumber\\
\amp\amp + \mathcal{O}\!\left(1\over \N^5\right)
\eea

\vspace{0.1cm}\begin{center}{\em f. Octic Order}\end{center}\vspace{0.2cm}

At octic order $\sim\phii^8$ we find 
\bea
(2\lambda)^2\langle\dt(\xv)\dt(\ov)\rangle_8\amp=\amp
{8(72\Delta^2+39\Delta^4+16\Delta^6+3\Delta^8)\over \N^4} \nonumber\\
\amp\amp +\mathcal{O}\!\left(1\over \N^5\right)
\eea

\section{Two-Point Function for Even Number of Fields}\label{EvenN}

When $\N$ is even the expression always involves the polylog function that we found for $\N=2$, plus corrections that depend on $\N$. We find that the form of the answer is
\bea
(2\lambda)^2\langle\dt(\xv)\dt(\ov)\rangle \amp=\amp \mbox{Li}_2(\Delta^2)+{P_\N(\Delta^2)\over\Delta^{\N-4}}\nonumber\\
\amp\amp+{\bar{P}_\N(\Delta^2)\ln(1-\Delta^2)\over\Delta^{\N-2}},\,\,\,\,\,\,\,\,\,\,\,
\eea
where 
\bea 
\mbox{ $P_\N(\Delta^2)$ is a polynomial of degree $(\N-4)/2$},\\
\mbox{ $\bar{P}_\N(\Delta^2)$ is a polynomial of degree $(\N-2)/2$}.
\eea
For $\N=4$, we find
\bea
P_4(\Delta^2)\amp=\amp-1,\\
\bar{P}_4(\Delta^2)\amp=\amp-1+\Delta^2.
\eea
For $\N=6$, we find
\bea
P_6(\Delta^2)\amp=\amp{1\over2}-{7\over4}\Delta^2,\\
\bar{P}_6(\Delta^2)\amp=\amp {1\over2}-2\Delta^2+{3\over2}\Delta^4.
\eea
For $\N=8$, we find
\bea
P_8(\Delta^2)\amp=\amp -{1\over3}+{4\over3}\Delta^2-{85\over36}\Delta^4,\\
\bar{P}_8(\Delta^2)\amp=\amp -{1\over3}+{3\over2}\Delta^2-3\Delta^4+{11\over6}\Delta^6.
\eea

\section{Alternative Derivation of Time-Delay Spectra}\label{Integral}

Here we write the two-point function as a multidimensional integral. 
It is convenient now to switch to a vector notation thus making the components of $\phi$ explicit. 
\bea
&&\tilde \phi(\ov,t) \equiv \vec \phi_x \equiv (X_1, X_2, X_3, \dots, X_\N), \\
&&\tilde \phi(\xv,t) \equiv \vec \phi_0 \equiv (X_{\N+1}, X_{\N+2}, X_{\N+3}, \dots, X_{2\N}),\,\,\,\,\,\,\,\,\,\,\,\,\,\, \\
&&\vec X \equiv (X_1, X_2, X_3, \dots, X_{2\N}).
    \label{3.1}
\eea
The average value of a function $F$ of a random variable $\vec X$  with
probability distribution function $p(X)$ is given by
\beq
\la F[X] \ra = \int dX p(X) F[X].
\eeq

Since we are using a free field approximation, $\vec X$ follows a joint Gaussian distribution
\beq
p(X) = {1\over (2\pi)^{\N} \!\sqrt {\det(\Sigma)}} \exp \left (-{1\over 2} X^T \Sigma^{-1} X \right)\!,\,\,\,\,\,
\eeq
where
\beq
\Sigma_{ij} = \la X_i X_j \ra 
\eeq
is the correlation matrix.
The components of $\Sigma$ can be easily calculated using the commutation relations for the creation and annihilation operators in $\phi(\xv, t)$, from Eq.~(\ref{eqn:phi(x)}). Due to the high degree of symmetry the matrix itself has a very simple structure: 
\beq
\Sigma_{ij}= \frac{1}{\N}(\delta_{i,j}+\delta_{i,(j\pm \N)}\Delta).
\eeq 

Finally, we can write the two-point function as:
\bea
\amp\amp\!\!\!\!\!\!\!\!\!(2\lambda)^2\la \delta t(\xv) \delta t(\ov) \ra \nonumber\\
\amp=\amp \int\! \frac{\N^2 d \vec X}{(2 \pi)^{\N} (1-\Delta^2)^{\frac{\N}{2}}}\log(|\vec \phi_0|^2)\log(|\vec \phi_x|^2)\times  \nonumber \\ 
\amp\amp\exp \{-\frac{\N}{2(1-\Delta^2)}[|\vec \phi_0|^2+|\vec \phi_x|^2 -2\N\Delta \vec \phi_0\! \cdot\! \vec \phi_x]\}.\,\,\,\,\,\,\,
\label{3.4} 
\eea
Then we can Taylor expand the integrand in powers of $\Delta$. We see that all odd powers of $\Delta$ vanish, leaving only even powers in the expansion. Performing these integrals term by term in the Taylor expansion, leads to the results reported earlier in eqs.~(\ref{FullExpansionTerms},\,\ref{FullExpansion},\,\ref{Coeff}).

\end{appendix}

\end{document}